# Optimal Sound-Absorbing Structures

*Min Yang[1], Shuyu Chen[3], Caixing Fu[1,3], Ping Sheng[1,2,\*]*

[1]*Department of Physics, HKUST, Clear Water Bay, Kowloon, Hong Kong, China*

[2]*Institute for Advanced Study, HKUST, Clear Water Bay, Kowloon, Hong Kong, China*

[3]*Acoustic Metamaterials Co. Ltd, No.2 Science Park West Avenue*

*Hong Kong Science Park, Shatin, New Territories, Hong Kong, China*

**Abstract**

Causal nature of the acoustic response, for any materials or structures, dictates an inequality that relates the absorption spectrum of the sample to its thickness. We present a general recipe for constructing sound-absorbing structures that can attain near-equality for the causal relation with very high absorption performance; such structures are denoted "optimal." Our strategy involves using carefully designed acoustic metamaterials as backing to a thin layer of conventional sound absorbing material, e.g., acoustic sponge. By using this design approach, we have realized a 12 cm-thick structure that exhibits broadband, near-perfect flat absorption spectrum starting at around 400 Hz. From the causal relation, the calculated minimum sample thickness is 11.5 cm for the observed absorption spectrum. We present the theory that underlies such absorption performance, involving the evanescent waves and their interaction with a dissipative medium, and show the excellent agreement with the experiment.

\*Corresponding author. Email: sheng@ust.hk

Sound absorption is important for room acoustics and remediation of noise that can arise from machines, rail cars, or cooling fans of computer server arrays in search engines or cloud computing. Traditional means of acoustic absorption, such as porous and fibrous *(1, 2)*, or gradient index materials *(3)*, can be either bulky or structurally weak. Micro-perforated panels with a tuned cavity depth behind the panels are effective for indoor sound absorption within a certain frequency range *(4-6)*, but not broadly applicable. During the past decade, locally resonant artificial structures *(7-13)*, acoustic metamaterials *(14, 15)*, and metasurfaces *(16, 17)* have shown diverse functionalities in the manipulation of sound such as negative refraction *(18-20)*, subwavelength imaging *(21-23)*, cloaking *(24, 25)*, one-way transmittance *(26, 27)*, and even highly efficient sound absorption within a compact volume *(28-38)*. However, due to the dispersive nature of resonances, these applications are narrowband in character. It is thus reasonable to ask whether there is a way to define the "best" absorber performance and its associated limitations. Defining the limit can inform us the potential that may still exist for better absorption performance, so as to induce efforts for improvement.

Material response functions for the electromagnetic and acoustic waves must satisfy the causality principle *(39)*. For the electromagnetic waves, the causal nature of the material response function was found to result in an inequality that relates a given absorption performance to the sample thickness *(40, 41)*. Adapted to acoustics, this relation (for sound waves propagating in air) can be expressed in the following form for a flat absorbing material (or structure) with thickness $d$ sitting on a reflecting substrate:

$$d \geq \frac{1}{4\pi^2} \frac{B_{eff}}{B_0} \left| \int_0^\infty \ln[1 - A(\lambda)] d\lambda \right|, \quad (1)$$

where $\lambda$ denotes the sound wavelength in air, $A(\lambda)$ is the absorption coefficient, $B_{eff}$ denotes the effective bulk modulus of the sound absorbing structure, and $B_0$ is the bulk modulus of air. A detailed derivation of Eq. (1) is given in the Supplementary Information. We define a sound absorbing structure to be optimal if equality or near-equality can be attained in the above relation. Some obvious implications immediately follow from Eq. (1). For example, total absorption within a finite frequency range is not possible for any finite thickness sample. Also, high absorption at low frequencies would dominate the contribution to sample thickness. However, $A(\lambda) \sim 1$ at a low frequency is entirely possible for a very subwavelength sample thickness *(29)*, provided the absorption peak is narrow.

While causal optimality by itself does not guarantee high absorption, it can be used in conjunction with the following two questions to assess the potential for improving the sample performance. The first is for a given $d$, what is the best high absorption performance one can achieve within a certain frequency range? The second is that for an objective $A(\lambda)$, what is the minimum sample thickness required? In this work we show that there is a general recipe for designing sound absorbing structures which can achieve near-equality in the causal relation Eq. (1) with a flat, near-perfect absorption spectrum starting at 400 Hz. In particular, the right hand side of Eq. (1) is shown to yield 11.5 cm, while the actual sample thickness is 12 cm. A similar optimal sample, with $d$=6 cm, was fabricated that displayed the same flat, near-perfect absorption spectrum above 800 Hz.

To motivate our design strategy, consider a layer of acoustic sponge with thickness $h$ sitting on a reflecting surface. Since a node must exist at the reflecting surface, high absorption becomes possible only when $\lambda_s/4 < h$, so that there is appreciable wave amplitude inside the sponge. Here $\lambda_s$ denotes the wavelength in sponge. In contrast, if the sponge is backed by a soft boundary, then an anti-node exists at the substrate boundary, and the absorption for $\lambda_s/4 > h$ becomes much better than that for the hard boundary, owing to the larger wave amplitude inside the sponge. However, soft boundary is by no means optimal. Near-perfect absorption would occur only when there is impedance matching so that there is no reflection; hence impedance matching with air is the goal.

To realize the optimal backing to a thin layer of sponge, we use a pre-designed acoustic metamaterial that is constrained by the objective $A(\lambda)$ which is ~1 above a low frequency cutoff. We recall that the impedance of a planar resonator at the resonance frequency is close to zero; hence it can approximate a soft boundary. For an acoustic metamaterial with multiple resonances, the total impedance can be expressed as *(29, 42)*

$$Z = \frac{i\rho_0 d}{\omega}\left(\sum_n \frac{\alpha_n}{\Omega_n^2 - \omega^2 - i\beta\omega}\right)^{-1}, \quad (2)$$

where $Z \equiv p/v$ denotes the surface impedance, $p$ being the pressure, $v$ being the displacement velocity in response to the pressure at the surface of the metamaterial, $\rho_0$ is the density of air, $\omega$ is the angular frequency, $\Omega_n$ denotes the $n$th resonance frequency, $\alpha_n$ is its dimensionless oscillator strength, and $\beta \ll \omega$ describes the weak system dissipation for the acoustic metamaterial. Definition of the oscillator strength is given below. For Eq. (2) to be an accurate description of the acoustic metamaterial, the lateral size of the metamaterial units, assuming them to be periodically arranged, must be very subwavelength in scale so that the diffraction effects can be neglected. This is the case in the present work.

We target impedance that is flat in frequency and close to the air impedance $Z_0 = \rho_0 v_0$ starting at a lower cutoff frequency $\Omega_1$. Here $v_0 = 343$ m/s denotes the speed of sound in air. In the idealized case where we have a continuum distribution of resonances, Eq. (2) can be converted to an integral,

$$Z \simeq \lim_{\beta \to 0} \frac{i\rho_0 d}{\omega} \left( \int_{\Omega_1}^{\infty} \frac{\alpha(\Omega)D(\Omega)}{\Omega^2 - \omega^2 - i\beta\omega} d\Omega \right)^{-1}, \quad (3)$$

where $D(\Omega)$ denotes the mode density per unit frequency. For the particular choice of $\alpha(\Omega)D(\Omega) = \mu$, a frequency independent constant, the integral can be integrated to yield

$$Z = \frac{2\rho_0 d / \mu}{\pi - 2i \tanh^{-1}(\Omega_1 / \omega)}. \quad (4)$$

By requiring $\mu = 2\rho_0 d / (\pi Z_0)$, the behavior for the real and imaginary parts of $Z$ is shown in Fig. 1A. The imaginary part, owing to the oscillatory nature of the integrand in Eq. (3), rapidly decays to zero for $\omega > \Omega_1$. However, the real part of the impedance is seen to approach the impedance matching condition $Z / Z_0 = 1$ beyond $\Omega_1$. The absorption spectrum can be calculated by using the formula

$$A = 1 - \left| \frac{(Z/Z_0) - 1}{(Z/Z_0) + 1} \right|^2 = 1 - \left| \frac{\tanh^{-1}(\Omega_1 / \omega)}{\pi - i \tanh^{-1}(\Omega_1 / \omega)} \right|^2. \quad (5)$$

This is plotted in Fig. 1B. In this idealized case, just the acoustic metamaterial alone can already achieve near-perfect broadband absorption, requiring only infinitesimal dissipation coefficient. By substituting $A(\lambda)$ as expressed by Eq. (5) into Eq. (1), with $\lambda = 2\pi v_0 / \omega$ and $B_{eff} = B_0$, we obtain $d \geq \lambda_1 / \pi^2$ with $\lambda_1 \equiv 2\pi v_0 / \Omega_1$. Hence for a cutoff frequency of $\Omega_1 / 2\pi = 312$ Hz, we have a minimum sample thickness of ~11.1 cm. The lesson from this idealized case is that flat impedance as a function of frequency requires the mode density to be inversely proportional to the oscillator strength, i.e., $\mu$ should be frequency independent. To achieve other target absorption spectrum, one should adjust $\mu(\Omega) = \alpha(\Omega)D(\Omega)$ so that the calculated absorption can best approximate the target spectrum.

For an acoustic metamaterial with a given cross sectional dimension, the modes are necessarily discrete and it is difficult to avoid a choppy absorption spectrum with large swings of high absorption followed by deep valleys. We show below that a thin layer of acoustic sponge in front of the acoustic metamaterial array can qualitatively alter the nature of absorption, leading to a flat absorption spectrum similar to the idealized case with a low frequency cutoff. Such an effect has been previously observed *(43)* in a one dimensional array of slit channels with a wire mesh placed on top. In the present case we offer a quantitative theory for such effect, as well as a general design strategy and the optimality perspective based on causality.

Consider an acoustic metamaterial sample whose cross section is a periodic two-dimensional square lattice with a periodicity of $L$=4.55 cm. A single unit is shown in Fig. 2A. While a variety of local

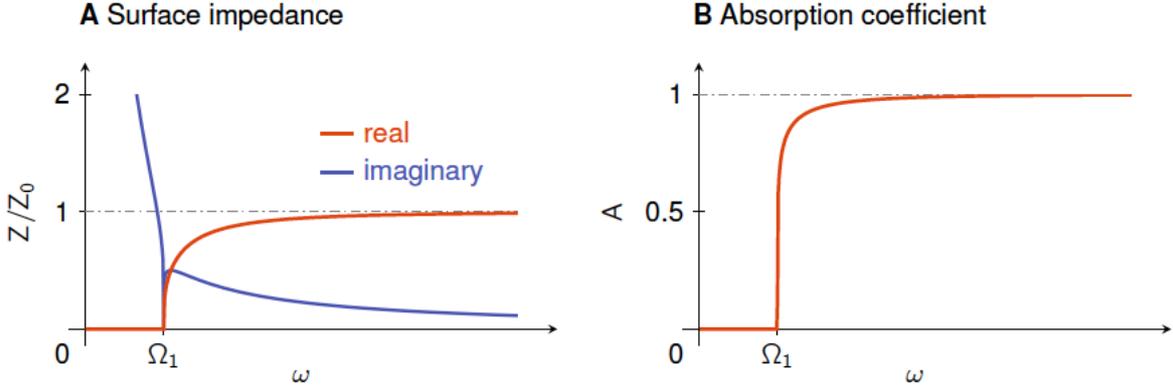

**Fig. 1 Broadband sound absorption and the optimally designed surface impedance.** Here we consider an idealized acoustic metamaterial with continuously distributed resonance frequencies above a cutoff $\Omega_1$. If the density of modes $D(\Omega)$ is inversely proportional to the oscillator strength $\alpha(\Omega)$ so that $\alpha(\Omega)D(\Omega) = 2\rho_0 d/(\pi Z_0)$, then the surface impedance $Z$ has the behavior as shown in (**A**), in which the real part quickly approaches the air impedance $Z_0$ above the cutoff $\Omega_1$, whereas the imaginary part decays towards zero. As the imaginary part of $Z$ contributes to the magnitude of reflection only in the form of $\left[\text{Im}(Z/Z_0)\right]^2$, hence its effect on absorption rapidly diminishes beyond $\Omega_1$. In this idealized case the acoustic metamaterial by itself can exhibit a near-perfect absorption spectrum above $\Omega_1$, as shown in (**B**). Substituting such an absorption spectrum into Eq. (1) leads to an inequality $d \geq d_{\min} = \lambda_1/\pi^2$, where $\lambda_1 \equiv 2\pi v_0/\Omega_1$. For $\Omega_1/2\pi = 312$ Hz and the absorption spectrum shown in (**B**), $d_{\min} = 11.1$ cm.

resonators, such as decorated membranes *(29)*, can be used to design the metamaterial unit, here we chose Fabry-Pérot (FP) resonators in which $B_{eff} = B_0$. Within each unit there are 16 FP channels each with a square cross section that is ~1.038 cm on the side, separated from each other with a 1 mm thick wall. The first order FP resonance frequency of the *n*th channel is given by $\Omega_n = \pi v_0/(2\ell_n)$, where $\ell_n$ denotes its channel length. For simplicity, we ignore the contributions of higher order FP resonances in the design process; their effect is mostly at higher frequencies and will be mentioned below. If we let *n*=1 to denote the longest channel so that $\Omega_1 = 2\pi \times 372.8$ Hz is the lowest resonance frequency, then the position of each channel in the unit is shown in Fig. 2A, where the (first order) resonance frequency of each channel increases with increasing *n*. We fold the longer channels in order to obtain a compact structure that can approach the thickness $\bar{d} = \sum_{n=1}^{16} \ell_n/16$. The folding, designed by computer simulations, can be seen for the blue- and pink-colored channels shown in Fig. 2A. In our particular case $\bar{d} = 10.58$ cm, whereas the actual sample, with folded channels, has a thickness *d*=11.06 cm. The relevant normalized eigenmode is given by $\phi_n(z) = \sqrt{4\Omega_n/\pi v_0}\cos[\pi z/(2\ell_n)]$, where *z*=0 defines the front surface of the metamaterial unit. In response to applied pressure, the $z$ component air displacement velocity $v_n$ at the mouth of the $n$th channel, $z=0$, is given by $v_n = -i\omega g_n p_n$, where $p_n$ denotes the local

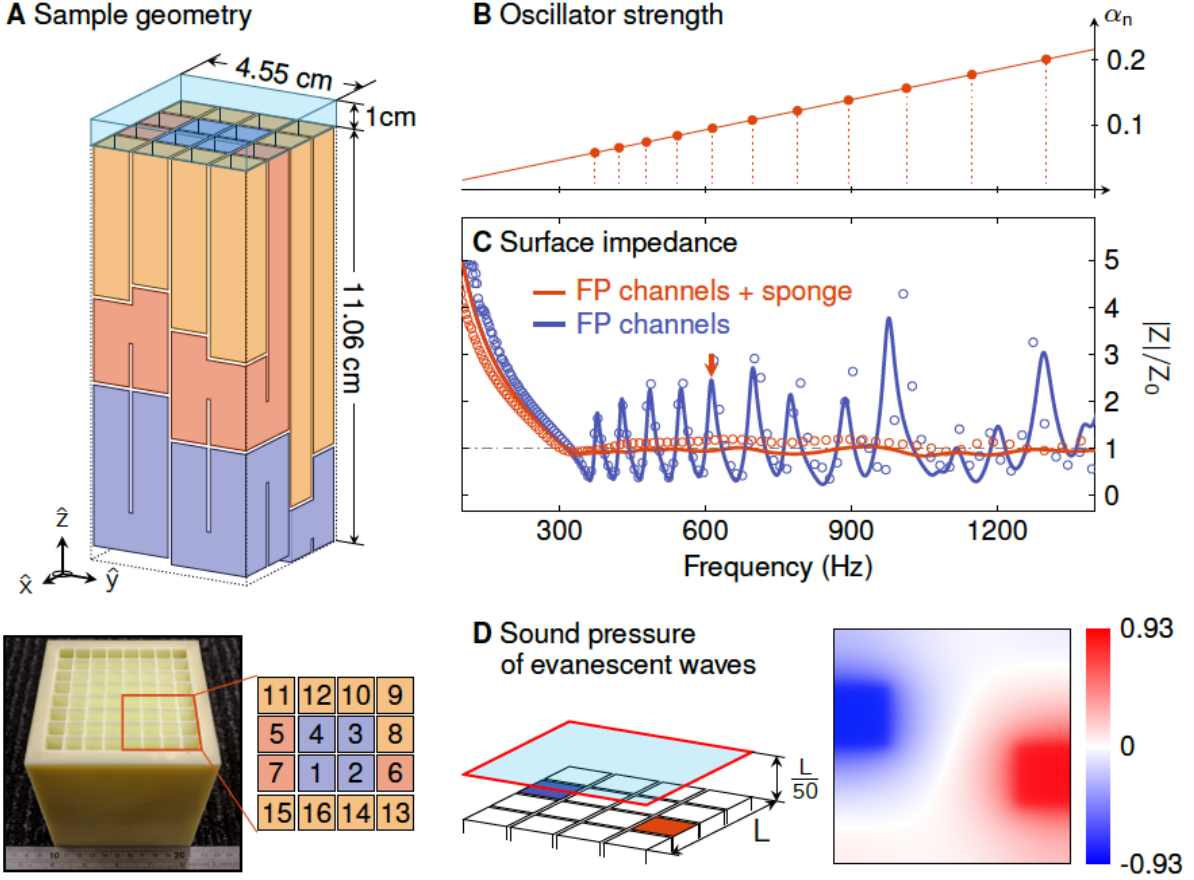

**Fig. 2 Metamaterial unit and its features**. (**A**) Schematics of the metamaterial unit consisting of 16 Fabry- Pérot (FP) channels, arrayed in a $4\times 4$ square lattice. The channel's number denotes its order in the sequence of decreasing lengths. Blue channels are coiled by 3 foldings, pink channels are coiled by 2 foldings, and the orange channels are straight. The transparent cyan block represents the sponge placed on the top of the channels' top surface. A photo image of the unit is shown in the lower left panel. (**B**) The oscillator strength $\alpha_n$ of the FP channels' $\lambda/4$ resonances, plotted up to 1400 Hz. The points indicate the first 11 resonant frequencies, designed so that the mode density is inversely proportional to the magnitude of $\alpha_n$. (**C**) The surface impedance of the metamaterial unit, $Z_{bare}$ (blue) and that with 1 cm thick sponge placed on top (red), both plotted as a functions of frequency. The circles are deduced from the measured reflection coefficient while the curves are the theory predictions based on Eqs. (10) and (11). The red arrow denotes the frequency at which the simulations were carried out on the sound pressure field, shown in the right panel below. (**D**) Right panel: the simulated sound pressure field at 0.91 mm above the front surface of the metamaterial unit at 610 Hz, which is the anti-resonance frequency between the FP resonances of the 5th and 6th channels (blue and red squares), shown on the left panel. The color is indicative of the pressure amplitude normalized to that of the incident wave. At anti-resonance, there is no coupling to the propagating modes; only evanescent mode exists as explained in the text.

pressure value, and $g_n$ is the Green function defined by *(42)*

$$g_n = \frac{|\phi_n(0)|^2 / \int_0^{\ell_n} |\phi_n(z)|^2 \rho(z)dz}{\Omega_n^2 - \omega^2 - i\beta\omega} \quad , \tag{6}$$

with $\rho(z)$ being the local mass density. Here we show only the first order FP resonance. In actual calculations all the higher orders are included. The accurate form of $g_n$ and its derivation are given in the Supplementary Information. The numerator of $g_n$ is proportional to the oscillator strength, $\alpha_n = \rho_0 d |\phi_n(0)|^2 / \left[16 \int_0^{\ell_n} |\phi_n(z)|^2 \rho(z) dz\right]$. Since in the present case $\rho(z) = \rho_0$, it follows $\alpha_n = d\Omega_n/(4\pi v_0)$, which is shown in Fig. 2B for the 11 of the 16 resonance frequencies obeying the rule $\Omega_n = \Omega_1 \exp[(n-1)/8]$, with $\Omega_{16} = 2431$ Hz. This rule is designed so that the mode density $D_n = (\partial \Omega_n / \partial n)^{-1} = 8\Omega_n^{-1}$, which leads directly to the desired $\mu = \alpha_n D_n = 2\rho_0 d /(\pi Z_0)$ as taught by the idealized case for achieving a constant impedance $Z_0$. Given the value of $\Omega_n$, the $n$th channel length is uniquely determined. The arrangement of the 16 channels within the metamaterial unit (shown in Fig. 2A) is optimized subject to the geometric requirements of channel folding.

We have fabricated the designed metamaterial unit by using 3D printing with UV sensitive epoxy. Reflection measurement from four units (a photo image of the measured sample is shown in Fig. 2A), arranged in a square and placed against a reflecting wall, was carried out by using an impedance tube with square cross section that is 9 cm on a side and 44.75 cm in length. Two microphone detectors were used to measure the amplitude and phase of the waves, from which the reflection coefficient $R = (Z/Z_0 - 1)/(Z/Z_0 + 1)$ as well as the impedance of the sample can be deduced. The impedance tube has a cutoff frequency between 1400 to 1500 Hz, beyond which the measured results are inaccurate. The measured impedance is shown by the blue circles in Fig. 2C. It is seen that they oscillate around $Z_0$ with peaks and valleys. This is expected, since we have 16 discrete resonances, and the impedance peaks can be associated with the anti-resonances that are in-between the neighboring resonances. In fact, by treating each FP channel to be independent from the others, the impedance of the unit, denoted by $Z_{bare}$, can be written as

$$Z_{bare} = \frac{i}{\omega}\left(\frac{1}{16}\sum_{n=1}^{16} g_n\right)^{-1}, \tag{7}$$

in analogy with Eq. (2). It is easily appreciated that $Z_{bare}$ displays oscillations in a similar fashion as the measured results. However, if a 1 cm layer of acoustic sponge is placed on top of the unit, then the impedance is shown by the red circles in Fig. 2C. It is seen that the oscillations almost completely vanish. Below we show that the origin of this effect is due to the evanescent waves that laterally couple the $v_n$'s at the mouths of different FP channels, and their interaction with a highly dissipative medium.

For frequencies much less than 7.5 kHz, we have $L \ll \lambda$. In this regime not only the angular effect of the incident wave would be minimal, but also the observed impedance of the unit should be the homogenized effective value. At $z=0$, the unit's surface has different local impedances, owing to the different FB channel length. It implies that the pressure should be most generally written as $p(\bm{x})$, where $\bm{x} = (x, y)$ denotes the lateral coordinate, so as to reflect this inhomogeneity. By writing $p(\bm{x}) = \bar{p} + \delta p(\bm{x})$, where $\bar{p}$ denotes the value of $p(\bm{x})$ averaged over the surface area of the unit, and $\delta p(\bm{x})$ represents whatever that is leftover. The $\delta p(\bm{x})$ component averages to zero over the surface area of the unit, and it can only couple to the evanescent waves that decay exponentially away from $z=0$. This is because from the dispersion relation we have $|\bm{k}_\parallel|^2 + k_\perp^2 = (2\pi/\lambda)^2$; and since the $\bm{k}_\parallel$ components from the Fourier transform of $\delta p(\bm{x})$ must be larger than $2\pi/\lambda$ (since $L \ll \lambda$), it follows that $k_\perp^2 < 0$, hence evanescent along the $z$ direction. In contrast, the $\bm{k}_\parallel$ components of $\bar{p}$ are peaked at $|\bm{k}_\parallel| = 0$, hence $\bar{p}$ couples to the propagating modes. In Fig. 2D, we illustrate the $\delta p(\bm{x})$ component at $z = 0.91$ mm and 610 Hz by plotting the full wave simulation results from a normally incident plane wave, carried out by using the commercial package COMSOL. This particular frequency (shown by the red arrow in Fig. 2C) is at the anti-resonance between the 5$^{th}$ and 6$^{th}$ channels' FP resonances, defined by $\bar{p} = 0$ so that there are only evanescent waves.

From the above discussion, we expect the impedance measured at far field to be given by $Z = \bar{p}/\bar{v}$, where $\bar{v} = \sum_{n=1}^{16} v_n / 16$. However, locally we must have

$$v_n = -i\omega g_n (\bar{p} + \delta p_n), \qquad (8)$$

where $\delta p_n$ represents the value of $\delta p(\bm{x})$ at the $n$th FP channel location. It is to be noted that even though $\sum_{n=1}^{16} \delta p_n = 0$, yet $\sum_{n=1}^{16} g_n \delta p_n \neq 0$. Therefore it is clear that the lateral inhomogeneity can contribute to the renormalization of the bare impedance. To complete the picture, we expand $\delta p(\bm{x})$ in terms of the normalized Fourier basis functions $f_{\bm{\alpha}} = \exp(i\bm{k}_{\bm{\alpha}} \cdot \bm{x})/L$, where $\bm{\alpha} = (\alpha_x, \alpha_y)$ is discretized by the condition that the integral of $f_{\bm{\alpha}}$ over the surface of the metamaterial unit must vanish. That means $|\bm{k}_{\bm{\alpha}}| = (2\pi/L)\sqrt{\alpha_x^2 + \alpha_y^2}$, with $\alpha_x, \alpha_y = \pm 1, \pm 2, \cdots$. Since $\delta p(\bm{x})$ is coupled to evanescent waves, each Fourier component must be associated with a $z$-variation given by $\exp\left[-\sqrt{|\mathbf{k}_\alpha|^2-(2\pi/\lambda)^2}\,z\right]$. That implies a nonzero $z$-derivative of $\delta p(\mathbf{x},z)$ which can couple to $v_n$'s. Through a simple derivation, given in the Supplementary Information, it can be shown that

$$\delta p_n = i\omega\rho_0 \sum_m \Lambda_{nm} v_m, \tag{9a}$$

$$\Lambda_{nm} = 16\sum_\alpha \frac{\sin^2(\alpha_x\pi/4)\sin^2(\alpha_y\pi/4)}{\pi^4 \alpha_x^2 \alpha_y^2 |\mathbf{k}_\alpha|} \exp[i\mathbf{k}_\alpha\bullet(\mathbf{x}_m-\mathbf{x}_n)], \tag{9b}$$

with $\mathbf{x}_n$ being the coordinate of the center of the $n$th channel at $z=0$, $\delta p_n = \delta p(\mathbf{x}_n)$, $v_m = v(\mathbf{x}_m)$, and $|\mathbf{k}_\alpha| \gg 2\pi/\lambda$. By substituting Eq. (9) into Eq. (8), we obtain

$$v_n = -i\omega\left(g_n + \omega^2 \rho_0 \sum_m g_n \Lambda_{nm} g_m + \cdots\right)\overline{p}$$

$$= -i\omega\left[g_n + \left(\omega^2\rho_0 g_n^2 \Lambda_{nn} + \omega^4 \rho_0^2 g_n^3 \Lambda_{nn}^2 + \cdots\right) + \omega^2\rho_0 \sum_{m\neq n} g_n \Lambda_{nm} g_m + \cdots\right]\overline{p}$$

$$= -i\omega\left(g_n + \frac{\omega^2\rho_0 g_n^2 \Lambda_{nn}}{1-\omega^2\rho_0 g_n \Lambda_{nn}} + \sum_m \Pi_{nm}\right)\overline{p}. \tag{10}$$

We have rearranged the series by separating out the terms involving $\Lambda_{nn}$, since $\Lambda_{nn} \gg \Lambda_{nm}$ by orders of magnitude. Numerically the last term in the bracket is also small and hence constitutes small adjustment in the result. By summing over $n$ on both sides, it is easily seen that $1/Z = 1/Z_{bare} + 1/\delta Z$ with

$$\delta Z = \frac{16i}{\omega}\left(\sum_n \frac{\omega^2\rho_0 g_n^2 \Lambda_{nn}}{1-\omega^2\rho_0 g_n \Lambda_{nn}} + \sum_n\sum_m \Pi_{nm}\right)^{-1}. \tag{11}$$

From Eq. (10), the evanescent wave effect is clearly seen in the appearance of the second term in the bracket. The resonances of the independent FP channels are identifiable by the frequencies at which $g_n$'s diverge. From Eq. (10) and the definition of the impedance, the renormalized resonance should be at the frequency which yields divergence of the coefficient in front of $\overline{p}$. If we add together the first two terms inside the bracket of Eq. (10), the result is a renormalized Green function given by $1/(g_n^{-1}-\omega^2\rho_0\Lambda_{nn})$; hence the resonance should occur at a frequency slightly below $\Omega_n$ where the real part of $g_n^{-1} = \omega^2\rho_0\Lambda_{nn}$. In Supplementary Information, this shift is calculated by using the exact expression for $g_n$ and compared with the experiment. In all the calculations we have used $\rho_0(1+i\beta/\omega)$ for the density of air, where $\beta = 14.2$ Hz serves to model the small air dissipation *(2)*. The solid blue line in Fig. 2C reflects the excellent agreement between the calculated and measured impedance.

In Fig. 3A, the measured absorption of the metamaterial unit is shown by blue circles. The solid curve is calculated by using Eq. (10) to evaluate the $Z$ and the reflection coefficient $R$, from which $A = 1 - |R|^2$. Very good agreement between theory and experiment is seen. Here the peaks in absorption are located at those frequencies where impedance matching is attained.

The absorption spectrum of the FP channels can be evaluated to be sub-optimal in character, even though the lower cutoff characteristic is clearly visible. To improve the absorption performance, we added 1 cm of sponge in front of the metamaterial unit. The separately measured absorption of a 1 cm thick sponge is also shown in Fig. 3A by the orange circles. To model the result, we treat the sponge as a uniform medium with a bulk modulus the same as that of air, justified by the fact that the sound predominantly travels through the pores. The fitted sponge mass density is given by $\rho_{sponge} = [1.4 + i(1420 \text{ Hz})/\omega]\rho_0$. Here the real part, $1.4\rho_0$, can be interpreted as due to the tortuosity of the pores, which lengthens the time of travel and hence an effectively lower sound speed. The imaginary part, which is two orders of magnitude larger than that for air, is due to the small size of the pores and the inevitably larger viscous boundary layer dissipation *(2)*.

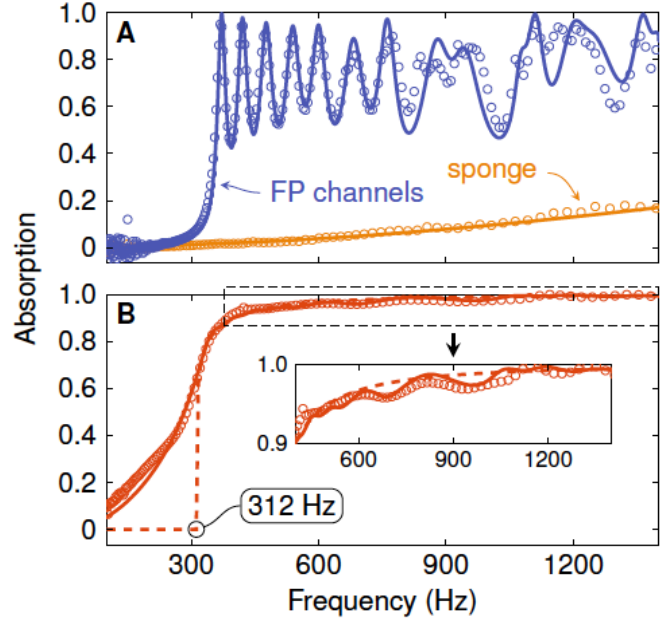

**Fig. 3 Experimental realization of broadband optimal sound absorption.** (**A**) The absorption spectrum of the acoustic metamaterial sample (blue circles), and that for a 1 cm sponge backed by a reflecting hard wall (orange circles). The blue and orange curves are calculated from theory with the parameter values given in the text. (**B**) The measured absorption coefficient (circles) of the sample, comprising the designed acoustic metamaterial covered by 1 cm of acoustic sponge. The solid curve is calculated from the theory in which the evanescent waves, in interaction with dissipative acoustic sponge, play a major role in filling the deep valleys in the absorption spectrum. For comparison, the spectrum for the idealized case, Fig. 1, is shown as dashed line. Near perfect absorption above the designed $\Omega_1 = 2\pi \times 312$ Hz can be seen, with excellent agreement between theory and experiment. To assess the optimality of the sample, we calculate the right hand side of Eq. (1) by using the experimental data below 1400 Hz and the simulation results beyond that. For high frequencies (i.e., beyond 1500 Hz) the higher order FP modes, in conjunction with the attendant shorter wavelength, guarantee excellent absorption by the acoustic sponge to be at least above the 97% range.

The significantly larger imaginary part of the sponge mass density has a dramatic effect on the real part of the impedance. This can be seen from Eqs. (10) and (11) by replacing $\rho_0$ by $\rho_{sponge}$, so that the total impedance of the combined sponge and metamaterial unit is given by

$$Z = \frac{i}{\omega}\left(G^{(e)}\right)^{-1} = \frac{i}{\omega}\left(\frac{1}{16}\sum_{n=1}^{16} g_n^{(e)}\right)^{-1}, \qquad (12)$$

where we have neglected the small contributions of $\Pi_{nm}$, and $\left(g_n^{(e)}\right)^{-1} = g_n^{-1} - \omega^2 \rho_{sponge} \Lambda_{nn}$. With the large imaginary part of $\rho_{sponge}$, at both the resonances and anti-resonances of the combined system the real part of $G^{(e)}$ would be zero, leaving only a nonzero imaginary part. As a result, there is a positive real part for $Z$ that is nearly flat as a function of frequency, owing to the designed mode density rule discussed earlier. This can be seen from the red curve in Fig. 2C, which was calculated from Eqs. (10) and (11) (the $\Pi_{nm}$ contributions included). It should be remarked that in order to remedy the effect of higher order FP resonances of the longer channels, which lowers the impedance at higher frequencies to a value slightly below $Z_0$, we have inserted 0.5 mm of sponge in channels 10-16. This has the positive smoothing effect on the impedance, without altering the designed resonance frequencies. In modeling, this added sponge is accounted for by increasing the dissipation coefficient $\beta$ of channels 10-16 by a factor of 3.3.

In Fig. 3B we show the absorption performance of the combined system, comprising 1 cm of sponge in front of our designed metamaterial unit. While the designed cutoff frequency is 312 Hz, there is an absorption tail below that, differing from the idealized case. Above the cutoff frequency the absorption reaches 90% at 400 Hz, then increases beyond that to an essentially flat, near-perfect absorption all the way to higher frequencies. The agreement between theory and experiment is excellent. Simulation results show that beyond 1400 Hz, the very high absorption is maintained with no sign of dropping; this is due to the fact that the shorter wavelength at higher frequencies, plus the plurality of higher order FP resonances (which makes the metamaterial unit's impedance low), insures the thin sponge layer absorption to be more effective than what is usually expected. By using the experimental absorption spectrum below 1400 Hz and the simulation results above that, the right hand side of Eq. (1) is evaluated to be 11.5 cm, whereas our sample thickness is 12 cm. However, if the channel folding can be improved so that the limit of $\bar{d}=10.58\,\text{cm}$ is reached, then the total thickness would be 11.58 cm, i.e., Eq. (1) essentially becomes equality.

A similarly optimal structure, with a metamaterial unit that has a 2.25 cm square cross section and a length of 5.5 cm, comprising 16 FP channels, has also been fabricated. Its absorption performance, with 5 mm of sponge in front (making the total sample thickness 6 cm), is very similar to that shown in Fig. 3B, except the horizontal scale is multiplied by two, i.e., now the 90% absorption starts at 800 Hz instead of 400 Hz. Experimental data of this smaller structure, together with the absorption performance of micro-perforated plate with a back cavity, also 6 cm in thickness, are shown and compared in the Supplementary Information.

We have conceived and implemented a general recipe for fabricating sound absorbing structures that display optimality as dictated by the causal nature of the acoustic response function, with unprecedented, near-perfect absorption spectrum starting at 400 Hz. Besides the broad applications potential (especially when the structures can be all metallic) of the design strategy that offers tunable absorption spectrum, perhaps the most important message of the present work is that metamaterials can be made to function in a true broadband fashion by utilizing the synergistic effect presented by evanescent waves, in interaction with a dissipative medium.

44. P.S. wishes to acknowledge Hong Kong Government's AoE/P-02/12 for the funding support of this research. We thank Feng Zhang for his help in sample characterizations at the beginning period of this project.

# Supplementary Information for

# Optimal Sound-Absorbing Structures

*Min Yang, Shuyu Chen, Caixing Fu, Ping Sheng*

## I. Causality constraint on sound absorbing structures

Consider a layer of composite material backed by a rigid reflective wall (Fig. S1A). At a given instant $t$, the reflected sound pressure $p_r(t)$ to an incident sound wave is a superposition of the direct reflection of the incoming sound pressure at this instant, $p_i(t)$, plus those in response to the incoming sound wave at earlier time, $p_i(t-\tau)$, with $\tau > 0$. Hence

$$p_r(t) = \int_0^\infty K(\tau) p_i(t-\tau) d\tau, \tag{S1}$$

where $K(\tau)$ is the response kernel in the time domain. By doing Fourier transform $p_{i/r}(\omega) = \int_{-\infty}^\infty p_{i/r}(t) e^{i\omega t} dt$, the reflection coefficient for each frequency may be expressed as

$$R(\omega) \equiv \frac{p_r(\omega)}{p_i(\omega)} = \int_0^\infty K(\tau) e^{i\omega\tau} d\tau. \tag{S2}$$

From Eq. (S2), $R(\omega)$ is an analytical function of complex $\omega$ in the upper half of the complex $\omega$ plane. In terms of the wavelength $\lambda = 2\pi v_0 / \omega$, where $v_0$ is the speed of sounds in air, that means $R(\lambda)$ has no singularities in the lower half-plane of complex $\lambda$, but may have zeros that represent total absorptions of incoming energy. Here the imaginary part of $\lambda$ reflects the dissipation.

To determine the constrain of reflection coefficient $R(\lambda)$ by the causality principle, we introduce an ancillary function $\tilde{R}(\lambda)$ after Fano and Rozanov *(1,2)*,

$$\tilde{R}(\lambda) \equiv R(\lambda) \prod_n \frac{\lambda - \lambda_n^*}{\lambda - \lambda_n}, \tag{S3}$$

where $\lambda_n$, satisfying $R(\lambda_n) = 0$, are the zeros located in the lower half-plane of complex $\lambda$, and $*$ stands for complex conjugation. Since $\tilde{R}$ has neither zeros nor poles at $\text{Im}(\lambda) < 0$, $\ell n \tilde{R}$ is an analytical function in the lower half-plane of complex $\lambda$ and the Cauchy theorem is valid, i.e., the integral over a closed contour $C$ in the lower half-plane of complex $\lambda$ should yield zero, where the contour consists of the real axis of and the semicircle $C_\infty$, which belongs to the lower half-plane and has infinite radius as shown in Fig. S1B. Hence

$$\int_C \ell n\tilde{R}d\lambda = \int_{-\infty}^{+\infty} \ell n\tilde{R}d\lambda + \int_{C_\infty} \ell n\tilde{R}d\lambda = 0. \tag{S4}$$

Note that $|\tilde{R}|=|R|$ at real wavelengths and $\ell n|R|$ is an even function of $\lambda$ according to its definition Eq. (S2). Taking the real part of Eq. (S4) yields

$$\text{Re}\int_C \ell n\tilde{R}d\lambda = 2\int_0^\infty \ell n|R|d\lambda + \text{Re}\int_{C_\infty} \ell nRd\lambda + \text{Re}\sum_n \int_{C_\infty} \ell n\frac{\lambda-\lambda_n^*}{\lambda-\lambda_n}d\lambda = 0. \tag{S5}$$

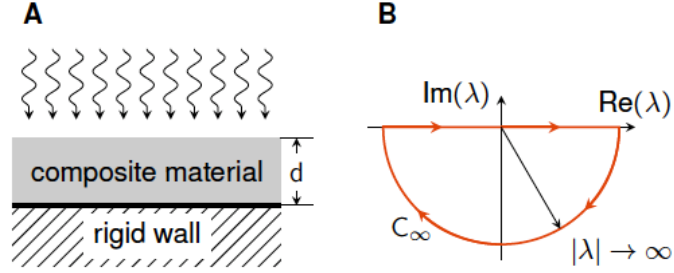

Fig. S1 (**A**) Schematic for the geometry of composite absorbing layer. (**B**) The contour for the integral in Eq. (S4).

To calculate the second integral on the right-hand-side of Eq. (S5), we consider the long-wavelength limit of $R$. At the long-wavelength limit, the composite material layer behaves like a uniform medium having an effective bulk modulus $B_{\text{eff}}$ (3). Its surface displacement $u$ under a pressure $p$ is therefore given by the relation (pressure)=(effective bulk modulus)*(strain), or $u = pd/B_{\text{eff}}$ with $d$ being the sample thickness. The resulting surface impedance is given by $Z = ip/(\omega u) = iZ_0 B_{\text{eff}} \lambda/(2\pi B_0 d)$ with $Z_0 = B_0/v_0$ being the air impedance and $B_0$ the bulk modulus of air. Therefore, the reflection coefficient $R = (Z-Z_0)/(Z+Z_0)$ is given by

$$R = \frac{1+i2\pi dB_0/(\lambda B_{\text{eff}})}{1-i2\pi dB_0/(\lambda B_{\text{eff}})}. \tag{S6}$$

Since $\lim_{|\lambda|\to\infty} \ell nR = i4\pi dB_0/(\lambda B_{\text{eff}})$, the contour integral is therefore given by

$$\int_{C_\infty} \ell nRd\lambda = \lim_{|\lambda|\to 0}\int_0^{-\pi} i\lambda \ell nRd\theta = 4\pi^2 dB_0/B_{\text{eff}}, \tag{S7}$$

where $\theta$ is the argument of complex $\lambda$. For the third integral on the right-hand-side of Eq. (S5), since $\lim_{|\lambda|\to\infty}\ell n[(\lambda-\lambda_n^*)/(\lambda-\lambda_n)] = i2\text{Im}(\lambda_n)/\lambda$, we have

$$\int_{C_\infty} \ell n\frac{\lambda-\lambda_n^*}{\lambda-\lambda_n}d\lambda = \lim_{|\lambda|\to\infty}\int_0^{-\pi} i\lambda \ell n\frac{\lambda-\lambda_n^*}{\lambda-\lambda_n}d\theta = 2\pi\text{Im}(\lambda_n). \tag{S8}$$

Substitution of Eqs. (S7) and (S8) into Eq. (S5) yields

$$-\int_0^\infty \ln|R(\lambda)|d\lambda = 2\pi^2 dB_0 / B_{\text{eff}} + \pi \sum_n \text{Im}(\lambda_n). \tag{S9}$$

As $[1-A(\lambda)] = |R(\lambda)|^2$, where $A(\lambda)$ stands for the absorption coefficient, and all $\lambda_n$ are in the lower half-plane, i.e., $\text{Im}(\lambda_n) < 0$, we therefore have the inequality

$$\left|\int_0^\infty \ln[1-A(\lambda)]d\lambda\right| = 2\left|\int_0^\infty \ln|R(\lambda)|d\lambda\right| \leq 4\pi^2 dB_0 / B_{\text{eff}}. \tag{S10}$$

II. **Derivation of pressure-velocity coupling equation with evanescent waves**

Since the surface impedance of the metamaterial unit is laterally inhomogeneous, it follows that the sound pressure field $p(\boldsymbol{x})$, where $\boldsymbol{x}$ denotes the lateral coordinate at the plane $z=0$, must necessarily be inhomogeneous as well. By decomposing the pressure field as $p(\boldsymbol{x}) = \bar{p} + \delta p(\boldsymbol{x})$, where $\bar{p}$ is the surface-averaged value, it has been shown in the main text that $\delta p(\boldsymbol{x})$ is only coupled to the evanescent waves that decay exponential away from $z=0$. In contrast, $\bar{p}$ couples to the far-field propagating modes. Therefore, the measured surface impedance should be given by $Z = \bar{p}/\bar{v}$ with $\bar{v} = d\bar{u}/dt$ being the surface-averaged $z$ component of the air displacement velocity. Reflection coefficient is given by $R = (Z-Z_0)/(Z+Z_0)$.

We expand $\delta p(\boldsymbol{x})$ in terms of the normalized Fourier basis function $f_{\boldsymbol{\alpha}}(\boldsymbol{x}) = \frac{1}{L}\exp[i\boldsymbol{k}_{\boldsymbol{\alpha}} \cdot \boldsymbol{x}]$, where $\boldsymbol{\alpha} = (\alpha_x, \alpha_y)$ is discretized by the condition that the area integral of $f_{\boldsymbol{\alpha}}$ over the surface of the metamaterial unit must vanish. That means $|\boldsymbol{k}_{\boldsymbol{\alpha}}| = (2\pi/L)\sqrt{\alpha_x^2 + \alpha_y^2}$, with $\alpha_x, \alpha_y = \pm 1, \pm 2, \cdots$:

$$\delta p(\boldsymbol{x}, z) = \sum_{\boldsymbol{\alpha}} \delta p_{\boldsymbol{\alpha}} f_{\boldsymbol{\alpha}}(\boldsymbol{x}) e^{-\sqrt{k_{\boldsymbol{\alpha}}^2 - k_0^2}\, z}, \tag{S11a}$$

where $\delta p_{\boldsymbol{\alpha}}$ denotes the expansion coefficient, and $k_0 = 2\pi/\lambda$. The exponential variation of $\delta p(\boldsymbol{x}, z)$ means that it can couple to the $z$ component of the air displacement velocity through Newton's law, $\partial \delta p / \partial z = -i\omega \rho_0 \delta v$, so that

$$\delta v(\boldsymbol{x}) = \delta v(\boldsymbol{x}, z=0) = \frac{-i}{\omega \rho_0} \sum_{\boldsymbol{\alpha}} \delta p_{\boldsymbol{\alpha}} \sqrt{|\boldsymbol{k}_{\boldsymbol{\alpha}}|^2 - k_0^2}\, f_{\boldsymbol{\alpha}}(\boldsymbol{x}). \tag{S11b}$$

By multiplying both sides of Eq. (S11b) by $f_{\boldsymbol{\alpha}'}^*(\boldsymbol{x})$ and integrating over the surface of the metamaterial unit's surface, we can solve for $\delta p_{\boldsymbol{\alpha}}$:

$$\delta p_\alpha = i\omega\rho_0 \frac{\int_{surface} v(\boldsymbol{x}) f_\alpha^*(\boldsymbol{x}) d\boldsymbol{x}}{\sqrt{|\boldsymbol{k}_\alpha|^2 - k_0^2}}, \tag{S11c}$$

where $v(\boldsymbol{x}) = \bar{v} + \delta v(\boldsymbol{x})$. It should be noted that in the above, the integral of $v(\boldsymbol{x}) f_\alpha^*(\boldsymbol{x})$ is the same as the integral of $\delta v(\boldsymbol{x}) f_\alpha^*(\boldsymbol{x})$, since the integral of $\bar{v} f_\alpha(\boldsymbol{x})$ is zero. By substituting Eq. (S11c) into Eq. (S11a) and then interchanging the order of summation and integration, we obtain

$$\delta p(\boldsymbol{x}) = \delta p(\boldsymbol{x}, z=0) = i\omega\rho_0 \int_{surface} \Lambda(\boldsymbol{x}, \boldsymbol{x}') v(\boldsymbol{x}') d\boldsymbol{x}', \tag{S12a}$$

where $\Lambda(\boldsymbol{x}, \boldsymbol{x}') \equiv \sum_\alpha f_\alpha^*(\boldsymbol{x}') f_\alpha(\boldsymbol{x}) / \sqrt{|\boldsymbol{k}_\alpha|^2 - k_0^2}$. Since $|\boldsymbol{k}_\alpha| \gg k_0$, we can approximate $\sqrt{|\boldsymbol{k}_\alpha|^2 - k_0^2}$ by $|\boldsymbol{k}_\alpha|$. By discretizing the 2D coordinate $\boldsymbol{x}$ by its 16 values, $\boldsymbol{x}_n$, that denotes the center position of the $n$th FP channel, and replacing $d\boldsymbol{x}'$ by $L^2/16$ and the integral by summation, we arrive at Eq. (9) in the main text:

$$\delta p_n = i\omega\rho_0 \sum_{m=1}^{16} \Lambda_{nm} v_m, \tag{S12b}$$

$$\Lambda_{nm} = \frac{16}{L^2} \frac{\int_{\sigma_n} f_\alpha(\mathbf{x}) d\mathbf{x} \int_{\sigma_m} f_\alpha^*(\mathbf{x}') d\mathbf{x}'}{|\boldsymbol{k}_\alpha|} = 16 \sum_\alpha \frac{\sin^2(\alpha_x \pi/4)\sin^2(\alpha_y \pi/4)}{\pi^4 \alpha_x^2 \alpha_y^2 |\boldsymbol{k}_\alpha|} \exp[i\boldsymbol{k}_\alpha \bullet (\mathbf{x}_m - \mathbf{x}_n)], \tag{S12c}$$

where $\sigma_n$ denotes the cross sectional area of the $n^{th}$ FP channel, and $v_m = v(\boldsymbol{x}_m)$, $\delta p_n = \delta p(\boldsymbol{x}_n)$.

**III. Shift of the resonance frequencies due to the renormalization effect by evanescent waves**

The total impedance of the metamaterial unit is given by

$$Z = \frac{1}{(1/Z_{bare}) + (1/\delta Z)}, \tag{S13a}$$

with

$$Z_{bare} = \frac{i}{\omega}\left(\frac{1}{16}\sum_{n=1}^{16} g_n\right)^{-1}, \tag{S13b}$$

$$\delta Z = \frac{16i}{\omega}\left(\sum_n \frac{\omega^2 \rho_0 g_n^2 \Lambda_{nn}}{1 - \omega^2 \rho_0 g_n \Lambda_{nn}} + \sum_{nm} \Pi_{nm}\right)^{-1}. \tag{S13c}$$

Here the accurate form of $g_n$, including all the higher order FP resonances for a channel of length $\ell_n$, can be obtained by the definition of $g_n = iv_n(0)/[\omega p_n(0)]$, with $p_n(z) = \cos\left[\omega(z+\ell_n)\sqrt{\rho_0/B_0}\right]$ and $v_n(z) = -i\sin\left[\omega(z+\ell_n)\sqrt{\rho_0/B_0}\right]/\sqrt{\rho_0 B_0}$. Therefore, $g_n$ is given by

$$g_n = \frac{\tan(\omega \ell_n \sqrt{\rho_0/B_0})}{\omega \sqrt{\rho_0 B_0}}. \tag{S14}$$

To account for the small dissipation presented by the air viscosity, one simply replace $\rho_0$ by $\rho = (1+i\beta/\omega)\rho_0$, in which the coefficient $\beta = 14.2$ Hz is an effective parameter characterizing the air's viscosity in FP channels and its value is obtained by fitting the experimental data. By expanding $g_n$ around its resonance frequency $\Omega_n = \pi v_0/(2\ell_n)$, we obtain the form given by Eq. (6) in the main text:

$$g_n \simeq 4\Omega_n \Big/ \left[\pi Z_0 \left(\Omega_n^2 - \omega^2 - i\beta\omega\right)\right].$$

Since the resonance modes are best detected by the imaginary part of the Green function, which in the present case is given by $\text{Im}(G) = \text{Im}(i/\omega Z)$, we have calculated $Z$ from Eq. (S13) and plotted the dimensionless quantity $\text{Im}(G)\Omega_1 Z_0$ in Fig. S2. Here $\Omega_1 = 2\pi \times 332$ Hz is the frequency of the 1st resonance mode.

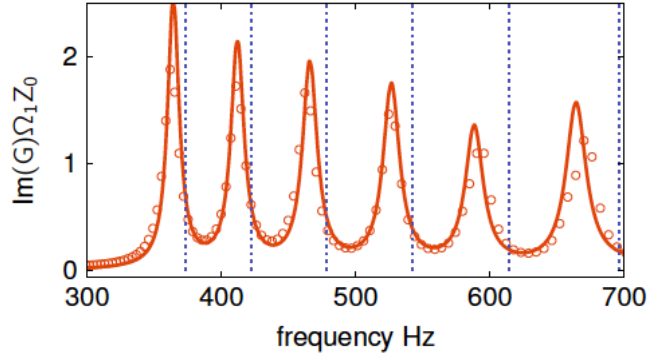

Fig. S2 The imaginary part of the Green function for the metamaterial unit, plotted as a function of frequency. The metamaterial unit consists of 16 FP channels in a $4 \times 4$ square lattice, shown in Fig. 2A of the main text. Here the solid curve is the prediction from the theory including the evanescent waves and the viscosity of air. The open circles are results deduced from experimental reflection measurements, where we have used the formula $\text{Im}(G)\Omega_1 Z_0 = \text{Im}\left\{-i\Omega_1(R-1)/[\omega(R+1)]\right\}$. It is seen that the peak positions, which indicate the renormalized resonances, are all down-shifted from the dotted blue lines that mark the original FP resonance frequencies ($\Omega_n$'s). Excellent theory-experiment agreement is seen.

As shown in Fig. S2, for the metamaterial unit the theoretically predicted positions of the newly emerged resonances (solid curve) fit the experiment (open circles) very well, and they all have a clear downward shift from the original FP resonances denoted by the vertical dotted lines. Physically, the downshift can be understood as due to the extra air mass participating in the resonant motion at the mouth of the FP channel, arising from the evanescent waves.

**IV. Comparison of micro-perforated panels with the broadband absorber**

As mentioned in the main text, causal optimality does not by itself guarantee broadband high acoustic absorption. In this section we give such an example: the micro-perforated panel (MPP) absorber *(4)*. The MPP has proven to be a good sound absorber, but only at multiple distinct frequencies, and such an absorption spectrum can also achieve optimality, provided the MPP's perforated hole diameter is small enough (see Fig. S3B).

Consider a panel that is $t = 0.2$ mm thick, with perforated holes that are arranged in a square lattice with a lattice constant $b = 2.5$ mm. The panel is backed by a chamber with $d = 6$ cm *(4)*. Maa's theoretical model *(4)* has proven to be very accurate for characterizing MPP's absorption. For uniform circular perforations with diameter $l$ the absorption is given by *(4)*

$$A(\omega) = \frac{4r}{(1+r)^2 + [\omega m - \cot(\omega d / v_0)^2]}. \tag{S15}$$

Here,

$$r = \frac{32\eta_0 t}{\sigma \rho_0 v_0 l} k_r, \quad k_r = \left[1 + \frac{k^2}{32}\right]^{1/2} + \frac{\sqrt{2}}{32} k \frac{l}{t},$$

$$m = \frac{t}{\sigma v_0} k_m, \quad k_m = 1 + \left[1 + \frac{k^2}{2}\right]^{-1/2} + 0.85 \frac{l}{t},$$

and

$$k = l\sqrt{\omega \rho_0 / (4\eta_0)}, \quad \sigma = l^2 \pi / (4b^2),$$

with $\eta_0$ being the viscosity coefficient of air, and the quantity $\sqrt{\eta_0 / (\omega \rho_0)}$ is usually denoted the viscous boundary layer thickness. Here $\eta_0$ is related to the effective air dissipation parameter $\beta$ in the main text, through the solution of the sound wave propagation in the FP channel.

In Fig. S3A, the solid blue line is the theory prediction of the MPP absorption with the parameter values given above, and $l = 0.2$ mm. The open circles are the experimental results *(4)*. The causal integral, Eq. (S10), gives $d_{min} \simeq d = 6$ cm, i.e., causal optimality is satisfied. In Fig. S3B, we plot the $d_{min}$ calculated from the predicted absorption spectra. It turns out that a critical perforation diameter $l_c \simeq 0.025$ mm exists, and for $l > l_c$ causal optimality is not satisfied. It is somewhat surprising that the critical value of the perforation hole diameter agrees so well with the "best" diameter of the holes as determined from an entirely different perspective *(4)*.

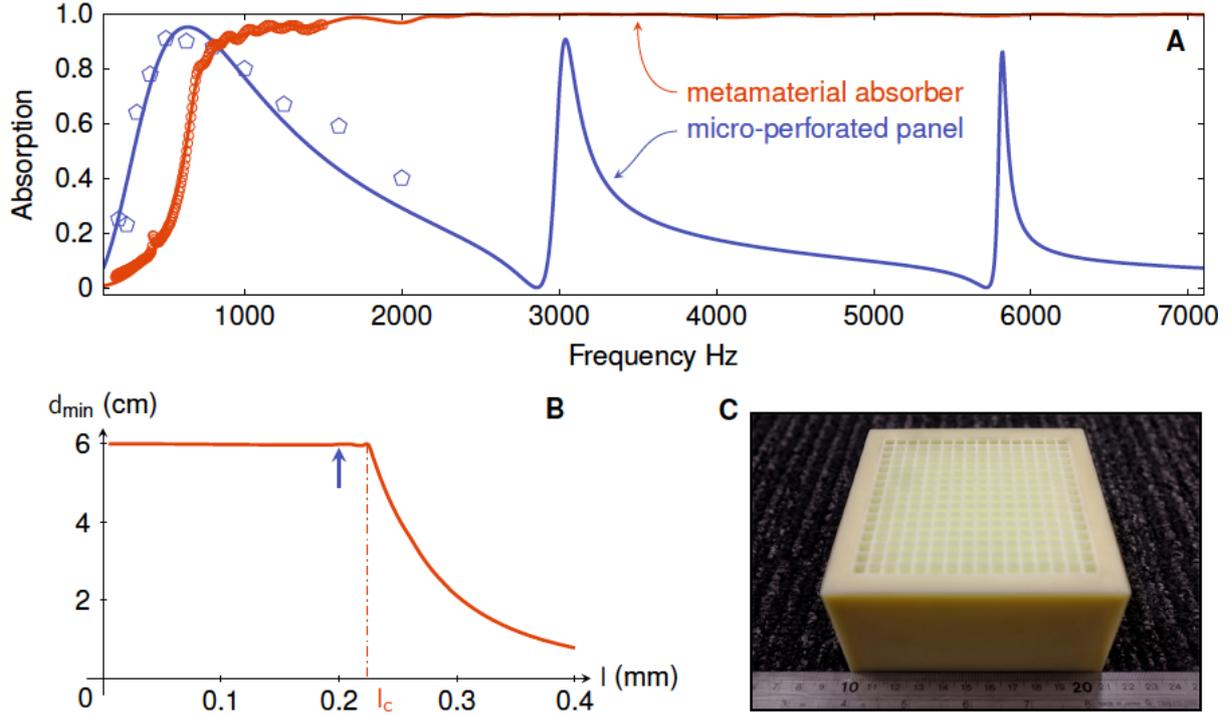

Fig. S3 (**A**) The absorption spectrums for the MPP (blue) with $l = 0.2$ mm and our broadband metamaterial absorber (red). The blue curve is from Maa's theoretical model Eq. (S15), and the pentagons are the experimental data from Maa's original paper *(4)*. The red curve is predicted by the theory presented in the previous section, and the red circles are the experimental data. (**B**) The minimal thickness $d_{min}$ determined by the causal integral of the absorption spectrum, Eq. (S10), for a MPP with diameters $l$, when $t = 0.2$ mm, $b = 2.5$ mm, and $d = 6$ cm. The blue arrow indicates the perforation's diameter of $l = 0.2$ mm, whose relevant absorption spectrum is shown in (**A**). (**C**) A photo image of the metamaterial absorber with the same thickness (6 cm) as the MPP.

To compare with the particular MPP absorber whose absorption spectrum is shown in Fig. S3A, we have redesigned our metamaterial absorber by scaling the structure in the main text to half the dimensions of the original, so that the metamaterial unit of this new absorber is a $2.275 \times 2.275 \times 5.5$ cm$^3$ cuboid. A photo image of the sample is shown in Fig. S3C. Sixteen such units were arranged into a square with the cross section that fit the cross section of the impedance tube (see Fig. S3C). The metamaterial unit was covered by a layer of 5 mm thick acoustic sponge, so that the total thickness of the absorber, 6 cm, is the same as that of the MPP in Maa's work *(4)*. The absorption spectrum of this metamaterial absorber is shown in Fig. S3A (red symbols for the experiment, red line the theory prediction). It is seen that near-perfect flat absorption starts around 800 Hz. The causal integral of this spectrum gives $d_{min} = 5.75$ cm, very close to the actual size of sample as well. It is clear that the two causally optimal structures exhibit absorption behaviors that are very different. The MPP starts its maximum absorption at a lower frequency, ~640 Hz, but quickly drops to nearly zero before its next resonance. This comparison emphasizes the fact that in the causal inequality, low frequency behavior dominates the contribution to the sample thickness. Here, the somewhat better low frequency absorption of the MPP is at the cost of degrading the absorption over large ranges of higher frequencies.